\documentclass[english,aps,prapplied,twocolumn,superscriptaddress,longbibliography]{revtex4-2}
\usepackage[urlcolor=blue, hyperindex, colorlinks, bookmarks=true]{hyperref}

\usepackage{graphicx}
\usepackage{dcolumn}
\usepackage{bm}
\usepackage{amsmath,xcolor,amssymb}
\usepackage[normalem]{ulem}
\usepackage{bm}
\usepackage{soul}
\usepackage[utf8]{inputenc}
\usepackage[T1]{fontenc}
\usepackage{mathptmx}
\usepackage{mathtools}
\usepackage{braket}
\usepackage{float}
\RequirePackage[normalem]{ulem} 

\definecolor{new}{rgb}{.38,.6,.38}
\definecolor{old}{rgb}{1,0,0}
\definecolor{off}{rgb}{0,0,0}

\newcommand{\UCLA}{\affiliation{Department of Physics and Astronomy, University of California -- Los Angeles, Los Angeles, California 90095, USA}}
\newcommand{\CQSE}{\affiliation{Center for Quantum Science and Engineering, University of California -- Los Angeles, Los Angeles, California 90095, USA}}
\newcommand{\ECE}{\affiliation{Department of Electrical and Computer Engineering, University of California, Los Angeles, California 90095, USA}}
\newcommand{\HRL}{\affiliation{HRL Laboratories, LLC, 3011 Malibu Canyon Road, Malibu, California 90265, USA}}
\newcommand{\princeton}{\affiliation{Department of Physics, Princeton University, Princeton, New Jersey~08544, USA}}

\begin{document}
\title{Microwave response of electrically driven spins in a three-qubit quantum processor}

\author{Tanner M. Janda}
\UCLA
\CQSE
\author{Heun Mo Yoo}  
\UCLA
\author{Connor Nasseraddin} 
\UCLA
\CQSE
\ECE
\author{Adam R. Mills}
\princeton
\author{Zhaoyi Joy Zheng}
\princeton
\author{Jason R. Petta}
\email{petta@physics.ucla.edu}
\UCLA
\CQSE
\ECE
\HRL

\begin{abstract}
In electric dipole spin resonance (EDSR), a single spin is electrically driven in the field gradient produced by a micromagnet. While EDSR has enabled high fidelity gate operations in many devices, there are reports of unexpected non-linearities in the Rabi frequency as a function of microwave drive amplitude. We carefully measure the response of Loss-DiVincenzo (LD) single spin qubits to resonant drives as well as simultaneous resonant and off-resonant drives, as would be encountered in a realistic quantum processor.  With the microwave amplitude carefully calibrated, we find that the Rabi frequency scales linearly with drive amplitude, even when all three spins are driven simultaneously. We also determine that heating-induced resonance frequency shifts from off-resonant drives are comparable to typical temporal drifts.  Our results indicate that the previously observed nonlinear response is not a general feature of LD spin qubits.
\end{abstract}

\maketitle

It is estimated that $10^5$ -- $10^8$ physical qubits will be required to implement quantum algorithms that have widespread applications \cite{Dalzell2025,gidney2025factor2048bitrsa,webster2026pinnaclearchitecturereducingcost}. Due to the similarity of their fabrication with conventional transistors, SiMOS and Si/SiGe spin qubits are promising platforms for large-scale quantum information processing \cite{Loss1998,Burkard2023RMP}.  There are many recent demonstrations of high fidelity operations at the single and two-qubit level \cite{Madzik2022,Reiner2024,Tanttu2024,Noiri2022,mills2022twoqubit,xue2022quantum,Weinstein2023}.  However, as the qubit count increases, new error mechanisms such as crosstalk tend to emerge and require mitigation \cite{Chen2014,Yan2018,fuentes2026sixqubit}.

The first demonstration of a spin qubit used the exchange interaction to control two-electron spin states in a double quantum dot \cite{Petta2005}. In order to coherently manipulate a single spin, a resonant ac magnetic field must be applied to drive Rabi oscillations \cite{Slichter1990}. Early implementations generated the ac magnetic field by passing an ac current through a stripline \cite{Koppens2006,Pla2012,Veldhorst2014}. All-electrical control of single spins was later achieved by electrically driving the electron in an intrinsic spin-orbit field or the gradient field produced by a micromagnet, a process termed electric dipole spin resonance (EDSR) \cite{Tokura2006,Nowack2007,Pioro2008,Takeda2016,Yoneda2018,Zajac2017CNot}. EDSR has since been demonstrated in larger quantum dot arrays \cite{Philips2022,Neyens2024}.

A recent experiment by Undseth~\textit{et al.} reported large nonlinearities (i.e.~$>$ 1 MHz deviations) in the spin qubit Rabi frequency with increasing microwave drive amplitude \cite{undseth2023nonlinear}. Similar effects were shown for both single qubit drives and the simultaneous drive of two qubits, suggestive of crosstalk between microwave-driven qubits. Temperature-dependent resonance frequencies have been reported, motivating higher temperature qubit operation to mediate microwave heating effects \cite{undseth2023hotter,wu2025simultaneoushighfidelitysinglequbitgates}. Theoretical models attributing the frequency shift to phonons greatly underestimate any non-linearity and the experimental observations are still poorly understood \cite{heinz2025phonon,undseth2023nonlinear,undseth2023hotter}.  Moreover, it is unclear if the results are general to microwave driven Loss-DiVincenzo spin qubits or the result of a unique device limitation.


In this Article, we carefully examine the behavior of a three qubit device that is controlled using EDSR.  We first measure the Rabi frequencies $f_{\rm R}$ of the qubits as a function of microwave drive amplitude $A_{\rm MW}$ and find they closely follow Rabi's equation. We next investigate the effects of an off-resonant pulse that precedes a Ramsey sequence to emulate sequential gate operations, as well as during the free evolution period of a Ramsey sequence to emulate simultaneous gate operations. In both cases we find that the qubit resonance frequencies shift by at most 100 kHz. Lastly, we search for signatures of crosstalk when all three qubits are driven simultaneously. From the single qubit response and precisely calibrated microwave drive amplitude, we are able to predict the Rabi frequencies of three simultaneously driven qubits.  Our data closely track the predicted response, suggesting there is negligible crosstalk.

\begin{figure*}[t]
	\centering
	\includegraphics[width=2\columnwidth]{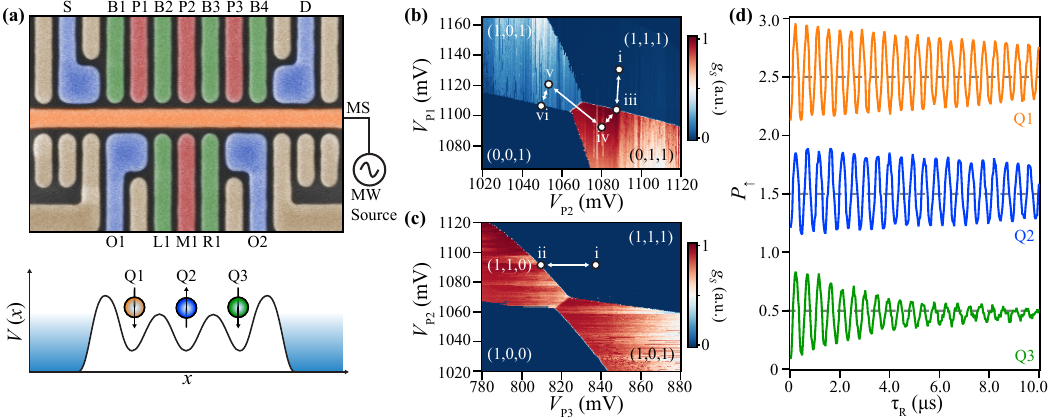}
	\caption{Device architecture and quantum control sequence. (a) Top:  False-color scanning electron microscope image of the TQD before the deposition of the Co micromagnet.  Bottom: Cartoon depicting the resulting electrostatic confinement potential $V(x)$.  (b--c) Charge stability diagrams acquired by measuring the charge sensor conductance $g_s$ as a function of gate voltages $V_{\rm Pi}$ and $V_{\rm Pj}$. The initialization, quantum control and readout sequence is indicated by the roman numerals (see text).  (d) Spin-up probability $P_\uparrow$ plotted as a function of microwave drive time $\tau_R$ for each of the three qubits.  Data are offset by 1.0 for clarity.}
	\label{fig:1}
\end{figure*}

Measurements are performed on an Intel ``Tunnel Falls'' triple quantum dot (TQD) \cite{George2024}.  A scanning electron microscope image of a similar device is shown in Fig.~\ref{fig:1}(a). Single electrons are confined beneath plunger gates P1 -- P3.  Barrier gates B2 and B3 control interdot tunnelling rates, while gates B1 and B4 control the tunneling rates to source (S) and drain (D) reservoirs.  The charge occupancy of the TQD is determined by measuring the conductance $g_{\rm s}$ of a charge sensor formed beneath gates L1, M1, and R1. The longitudinal and transverse magnetic field gradients generated by a cobalt micromagnet allow each spin qubit in the TQD to be selectively driven \cite{Tokura2006}.  At an applied magnetic field $B_{z}=337$ mT, each electron forms a Loss-DiVincenzo single-spin qubit with a resonance frequency in the range of 11.5 -- 11.7 GHz \cite{Loss1998}.

Figures~\ref{fig:1}(b,c) show pairwise charge stability diagrams with TQD charge states denoted by ($N_1$,~$N_2$,~$N_3$), where $N_i$ is the number of electrons in dot $i$. The initialization, quantum control, and readout sequence is overlaid on the charge stability diagrams with specific steps indicated by the roman numerals.  The initial spin state $\ket{\downarrow, \downarrow, \downarrow}$ is prepared through a combination of spin-selective tunneling to the charge reservoirs on either side of the TQD and shuttling \cite{Elzerman2004}. With the spins initialized at position (i) in Figs.~\ref{fig:1}(b,c), Rabi oscillations are driven by applying microwaves to gate MS. Qubit 3 (Q3) is read out via energy-dependent tunneling to the drain reservoir (i $\rightarrow$ ii), followed by Q1 readout at the source reservoir (ii $\rightarrow$ i $\rightarrow$ iii). Finally, we shuttle Q2 from dot 2 to dot 1 (iv $\rightarrow$ v), and then read out its spin state (v $\rightarrow$ vi). After traversing back to (1,1,1) by following the same sequence in reverse, the TQD is re-initialized in $\ket{\downarrow, \downarrow, \downarrow}$.  Figure~\ref{fig:1}(d) shows the spin-up probability $P_\uparrow$ as a function of Rabi drive time $\tau_{\rm R}$, demonstrating coherent control of each spin qubit.

\begin{figure}[b]
	\centering
	\includegraphics[width=\columnwidth]{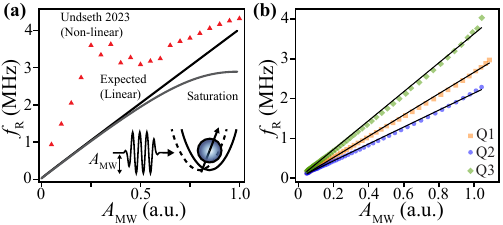}
	\caption{Linearity of single spin Rabi oscillations with drive amplitude. (a) Rabi's equation predicts a linear dependence of the Rabi frequency $f_{\rm R}$ on the microwave drive amplitude $A_{\rm MW}$. The non-linear behavior observed by Undseth \textit{et al.}~is shown for comparison \cite{undseth2023nonlinear}. Experimental limitations, such as the saturation of amplifiers at high powers, could also lead to slight deviations from linearity.  (b) Measured $f_\text{R}$ as a function of $A_{\rm MW}$ for each qubit, showing the expected linear dependence. Solid lines are linear fits to the data.}
	\label{fig:2}
\end{figure}

Conventional electron and nuclear spin resonance Rabi frequencies $f_{\rm R}$ scale linearly with the microwave drive amplitude $A_{\rm MW}$ \cite{Abragam1961,Slichter1990}.  An EDSR drive shifts the orbital position of the electron in a magnetic field gradient, resulting in an oscillating transverse magnetic field [see inset of Fig.~\ref{fig:2}(a)].  EDSR theories also predict a linear scaling of $f_{\rm R}$ with $A_{\rm MW}$ \cite{Laird2007,Tokura2006,Golovach2006}. In experiments, the saturation of microwave sources and mixers, often called ``compression,'' may give rise to a slight deviation from linear behavior at high drive amplitudes. The highly non-linear dependence observed by Undseth \textit{et al.}~is plotted along with the expected linear and saturated responses in Fig.~2(a) \cite{undseth2023nonlinear}.

We first characterize the response of the TQD in the simplest regime where only one spin qubit is driven at a time. In Fig.~\ref{fig:2}(b), $f_{\rm R}$ is plotted as a function of $A_\text{MW}$ for each qubit, where $A_\text{MW}$ is calibrated using a spectrum analyzer (see Appendix \ref{app:a}). All three qubits  exhibit the expected scaling $f_{\rm R} \propto A_{\rm MW}$, with linear fits to the data yielding $R^2>0.99$.  To provide a point of comparison, plots of $f_{\rm R}$ as a function of uncalibrated microwave amplitude are included in Appendix \ref{app:b} and do exhibit slight saturation.  These datasets highlight the importance of carefully calibrating the microwave drive circuit output.

\begin{figure}[t!]
	\centering
	\includegraphics[width=\columnwidth]{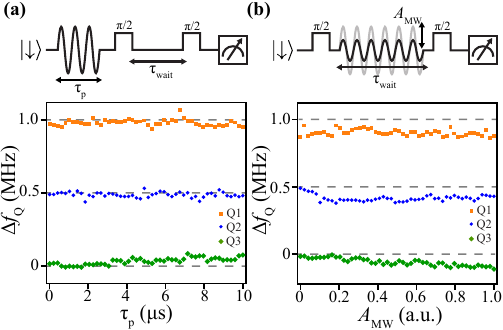}
	\caption{Sensitivity of qubit resonance frequencies to off-resonant drives. (a) 
    Qubit resonance frequencies are determined using a Ramsey pulse sequence. To assess the sensitivity of a qubit to off-resonant drives that will be present in multi-qubit implementations, the Ramsey sequence is preceded by a pre-pulse of duration $\tau_\text{p}$. The resulting qubit frequency shift $\Delta f_{\rm Q}$ is plotted as a function of $\tau_\text{p}$. (b) Ramsey pulse sequence with an off-resonant pulse applied during the entire free-evolution interval $\tau_\text{wait}$. $\Delta f_{\rm Q}$ is plotted as a function of $A_\text{\rm MW}$. Data are offset by 0.5 for clarity.}
	\label{fig:3}
\end{figure}

To perform quantum algorithms, microwave pulses must be applied to the qubits with varying frequencies and amplitudes. In Fig.~\ref{fig:3}(a), we investigate the impact of off-resonant pulses on the qubit resonant frequencies when applied before a Ramsey sequence. By changing either $A_{\text{MW}}$ or the duration $\tau_\text{p}$ of the pre-pulse, we emulate an increase in the number of gates applied prior to measurement \cite{undseth2023hotter}. We apply an off-resonant pulse with frequency $f_{\text{off}}=11.731$ GHz, which is detuned by 30 MHz from the nearest qubit frequency to avoid driving any of the qubits on resonance.

After applying the pre-pulse, we implement a Ramsey pulse sequence. We first execute a $\pi/2$ rotation 1 MHz detuned from the target qubit frequency and let the system freely evolve for a time interval $\tau_{\rm wait}$.  A second $\pi/2$ rotation is applied before measuring the spin-up probability $P_{\uparrow}$. By sweeping $\tau_\text{wait}$ we observe oscillations in $P_{\uparrow}$. The qubit resonance frequency can be extracted from the frequency of the Ramsey fringes (see Appendix \ref{app:c}). Figure~\ref{fig:3}(a) shows the outcome of our measurements, where the resulting qubit frequency shift $\Delta f_{\rm Q}$ is plotted as a function of 
$\tau_{\rm P}$.  We find that after applying an off-resonant pre-pulse for up to 10 $\mu$s, which is much longer than typical Rabi pulse lengths of 300 ns \cite{mills2022twoqubit}, our qubit resonance frequencies shift by no more than $\Delta f_{\rm Q}^{\rm max}$ = 50 kHz.  Measurements of typical qubit resonance frequency shifts over a 2 hour period are comparable in magnitude (see Appendix \ref{app:c}).

Sequential gate operations limit the performance of quantum algorithms as idling qubits dephase and accumulate errors \cite{fuentes2026sixqubit}. Quantum circuits can be optimized by applying gate operations in parallel, thereby reducing idle times.  To evaluate resonance frequency shifts during parallel gate operations, we implement a similar Ramsey sequence, but with the off-resonant pulse applied during the free-evolution interval [see Fig.~\ref{fig:3}(b)]. Our data again exhibit frequency shifts that are comparable with typical temporal drifts.   The initial frequency offset for Q1 is attributed to temporal frequency drift during the overnight data run.  In summary, the qubit resonance frequency shifts observed with pre-pulsing and simultaneous pulsing is comparable with typical temporal drifts (50 -- 100 kHz).  No significant or reproducible frequency shifts are observed in the data.

\begin{figure*}[t!]
	\centering
	\includegraphics[width=2\columnwidth]{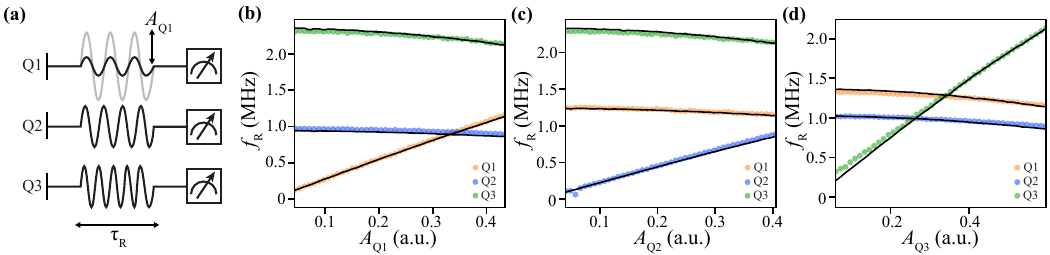}
	\caption{Linearity of three simultaneously driven single spin Rabi oscillations with drive power. (a) The three qubit drive tones are multiplexed into a single $IQ$ channel that modulates a vector microwave source driving the MS gate. We vary the drive amplitude $A_{\rm Qi}$ of a single qubit while holding the drive amplitude of the other two qubits fixed. (b -- d) Extracted $f_{\rm R}$ plotted as a function of $A_{\rm Qi}$. As the power delivered to one qubit is increased, we see a slight decrease in the Rabi frequency of the other two qubits. Solid lines are the measured power output by the microwave source scaled by the single qubit linear fits, which suggests that the curvature comes from microwave power saturation rather than qubit crosstalk.}
	\label{fig:4}
\end{figure*}

An earlier experiment reported non-linearities in the Rabi frequencies of simultaneously driven qubits, posing a difficult challenge for extending spin qubits towards complex quantum circuits \cite{undseth2023nonlinear}. As a final demonstration, in Fig.~\ref{fig:4} we investigate frequency shifts resulting from the simultaneous drive of all three spin qubits. In these measurements, we frequency multiplex three microwave drive tones and apply them to the shared MS gate (see Appendix \ref{app:a}).  By adjusting the amplitude of one drive tone (e.g.~$A_{\rm Q1}$) while keeping the other two constant, we can search for signatures of crosstalk [see Fig.~\ref{fig:4}(a)]. Varying the Rabi pulse length $\tau_{\rm R}$ reveals Rabi oscillations that are then fit to obtain Rabi frequencies. Similar to the single qubit case, we expect Rabi frequency to scale linearly with drive amplitude.  For example, when driving Q1 with varying $A_{\rm Q1}$ and holding $A_{\rm Q2}$ and $A_{\rm Q3}$ fixed, we expect $f_{\rm R}$ to scale linearly with $A_{\rm Q1}$ and the Rabi frequencies of the other two qubits to remain unchanged.

Our data are plotted in Figs.~\ref{fig:4}(b--d). We see that the Rabi frequency of the control qubit (the qubit driven with varying amplitude) scales linearly with that drive amplitude. We also observe a slight decrease in the Rabi frequencies of the two other qubits, which we expected to stay constant.  To more completely model the data, we measured the actual microwave source output amplitude for the exact parameters used in
Figs.~\ref{fig:4}(b--d) and scaled these values by the linear fits shown in Fig.~\ref{fig:2}(b) (see Appendix \ref{app:b}). The results of our analysis are plotted as the black solid lines in Figs.~\ref{fig:4}(b -- d), showing a strong correlation between the measured Rabi frequencies. All of the curvature and downward trends observed in the data can thus be explained by microwave compression in room temperature electronics, not qubit crosstalk.


In conclusion, we investigated the microwave response of an Intel TQD device that was driven using EDSR.  When driving a single qubit, we find that the Rabi frequencies are linear as a function of microwave drive amplitude.  In comparison with typical temporal resonance frequency drifts, the qubit Rabi frequencies are insensitive to pre-pulses applied before a Ramsey sequence and off-resonant pulses applied during the free-evolution interval of a Ramsey sequence.  Lastly, simultaneously driven Rabi oscillations for all three qubits are consistent with measurements factoring in the compression of the output of the microwave vector source.  Our results indicate that the previously reported strong non-linearities are not a general feature of LD spin qubits \cite{undseth2023nonlinear} and that industrially fabricated LD spin qubits have the potential to be scaled to larger system sizes.

\begin{acknowledgments}
Research was sponsored by the Army Research Office and was accomplished under
Cooperative Agreement No.~W911NF-22-2-0037 and Grant No.~W911NF-23-1-0104. The views and conclusions contained in this document are those of the authors and should not be interpreted as representing the official policies, either expressed or implied, of the Army Research Office or the U.S. Government. The U.S. Government is authorized to reproduce and distribute reprints for Government purposes notwithstanding any copyright notation herein.  We acknowledge support from Intel Corporation for providing the device, as well as technical conversations with Nathan Bishop, Joelle Corrigan, Matthew Curry, and Rene Otten.
\end{acknowledgments}


\appendix




\section{Experimental setup}
\label{app:a}

\begin{figure}[]
	\centering
	\includegraphics[width=\columnwidth]{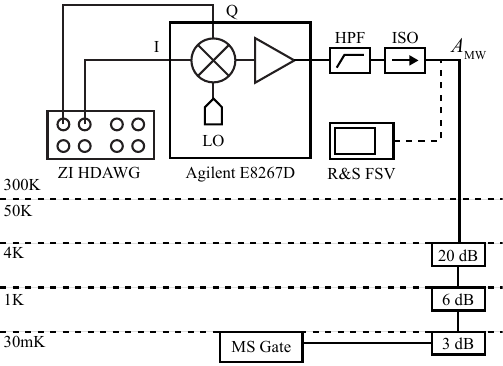}
	\caption{Block diagram depicting microwave signal generation and delivery to the device. An arbitrary waveform generator drives the $IQ$ ports of a vector microwave source, modulating the local oscillator $LO$. The modulated signal is high-pass filtered (HPF), isolated (ISO), and routed through several stages of cryogenic attenuation before it reaches the device. The output can optionally (dashed line) be connected to a spectrum analyzer to calibrate the output amplitude.}
	\label{sup:exp}
\end{figure}

The experimental setup used to generate microwaves is shown in Fig.~\ref{sup:exp}. We use a Zurich Instruments HDAWG to pulse plunger gates P1 -- P3 to navigate charge stability space. The HDAWG also produces pulses that modulate the $IQ$ ports of an Agilent E8267D vector signal generator. Pulse envelopes are defined by a raised cosine function to avoid spectral components that might drive other qubits \cite{xue2022quantum}. We vary the output power by either changing the amplification from the microwave source or by adjusting the $IQ$ pulse amplitude. We calibrate the power delivered to the top of the cryostat using a Rohde \& Schwarz FSV Spectrum Analyzer. A 10 dB attenuator is used during the calibration to avoid additional saturation effects coming from the FSV. In the main text figures, the overall scale $A_{\text{MW}}$~=~1 corresponds to an applied power of 30~dBm at the top of the cryostat. At this power, we observe Rabi oscillations that agree with the simulated transverse magnetic field gradient $0.4 - 0.5~\rm mT/nm$ \cite{Neyens2024}.

\section{Calculated Rabi frequency}
\label{app:b}

The qubits are driven on resonance for a time $\tau_{\rm R}$, revealing Rabi oscillations that can be fit to extract $f_{\rm R}$, as shown in Fig.~\ref{sup:power}. We calibrate $A_{\rm MW}$ using spectrum analyzer measurements described in Appendix \ref{app:a} to plot the data shown in Fig.~\ref{fig:2}(b). In Figs.~\ref{fig:4}(b--d), we calculate predicted Rabi frequencies, shown by black lines, based on single qubit data and room temperature power measurements. Modeling the Rabi frequency simply as a linear function of drive amplitude, $f_{R}~=~\alpha_i A_{\rm Qi}$,  we can fit $\alpha_i$ for $i=1,2,3$ corresponding to each qubit in the device using the single qubit data from Fig.~\ref{fig:2}(b). We measure the drive amplitude $A_{\rm Qi}$ of each qubit frequency for each experiment in Figs.~\ref{fig:4}(b--d). Then, using the linear fits from Fig.~\ref{fig:2}(b), we map the measured amplitude to a Rabi frequency. This gives us a prediction for the Rabi frequency of each qubit during simultaneous operation only using single qubit data.

\begin{figure}[h]
	\centering
	\includegraphics[width=\columnwidth]{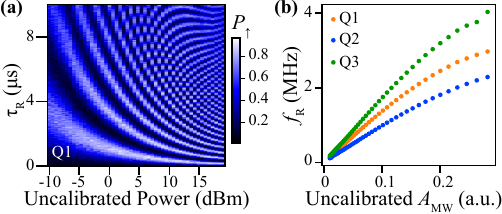}
	\caption{(a) Rabi oscillations of Q1 as a function of uncalibrated power. (b) Extracted $f_{\rm R}$ as a function of uncalibrated $A_{\rm MW}$, showing the saturation effect at high amplitudes due to compression of the microwave source.}
	\label{sup:power}
\end{figure}

\section{Extracting qubit frequencies}
\label{app:c}

\begin{figure}[]
	\centering
	\includegraphics[width=\columnwidth]{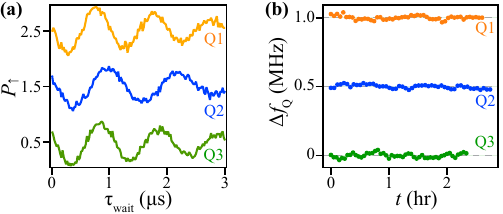}
	\caption{(a) Ramsey oscillations as a function of free evolution time $\tau_{\rm wait}$. Data are offset by 1.0 for clarity. (b) Qubit frequency shifts with time. Data are offset by 0.5 for clarity.}
	\label{sup:ramsey}
\end{figure}

Qubit frequency shifts are determined by performing a Ramsey experiment with single qubit drives detuned by $\delta$ from the qubit resonance frequency. The resulting readout probability $P_\uparrow$ oscillates as a function of free evolution time, $\tau_{\rm wait}$, as shown in Fig.~\ref{sup:ramsey}(a). Sweeping $\delta$ allows for a precise measurement of $f_{\rm Q}$. However, by fixing $\delta\neq0$, we have enough information to determine the qubit frequency shift $\Delta f_{\rm Q}$ while only performing a one dimensional sweep of $\tau_{\rm wait}$. For data shown in Fig.~\ref{fig:3} and Fig.~\ref{sup:ramsey}, we use a $\delta$~=~1 MHz for each qubit and fit the resulting oscillations to an exponentially damped sinusoidal function to extract $\Delta f_{\rm Q}$.

In Fig.~\ref{sup:ramsey}(b), we measure the temporal frequency shift of each qubit over a period of 2 hours. We do so by repeating a Ramsey sequence to match the data count shown in Fig.~\ref{fig:3}. We find that the maximum qubit frequency drift $\Delta f_{\rm Q}^{\rm max}$~=~50~--~75~kHz in this time frame, which is comparable to the frequency shifts observed during pre-pulsing and simultaneous drive experiments discussed in the main text. 

\bibliography{RMP_master2_bib_v4}

\end{document}